\documentclass[a4paper,11pt]{article}
\pdfoutput=1 

\usepackage{jcappub} 

\usepackage[T1]{fontenc} 
\usepackage{xcolor}

\title{
Quantum Creation of a Toy Universe without Inflation
}

\author{Yi Wang}
\author{and Mian Zhu}

\affiliation{Department of Physics, The Hong Kong University of Science and Technology, Clear Water Bay, Hong Kong S.A.R., P.R.China}
\affiliation{HKUST Jockey Club Institute for Advanced Study, The Hong Kong University of Science and Technology, Clear Water Bay, Hong Kong S.A.R., P.R.China}
\emailAdd{phyw@ust.hk}
\emailAdd{mzhuan@connect.ust.hk}
 
\abstract{
	We propose a toy model for the origin of the universe, where the scale-invariant fluctuations are generated together with the quantum creation process of the universe. The fluctuations arise inside an instanton in the Euclidean domain of time. In the Lorentzian point of view, the universe emerges with passive, coherent and scale-invariant fluctuations present from the beginning, without the need of inflation or a bounce. For this mechanism to work, we need anisotropic scaling in space and time, which is realized in a toy model of Horava-Lifshitz gravity with a Lifshitz scalar field. 
}

\begin{document}
	\maketitle
	\flushbottom
	
	\section{Introduction}
	\label{sec:intro}
	
	Inflation \cite{Guth:1980zm,Linde:1981mu,Albrecht:1982wi, Hawking:1981fz,Fang:1980wi,Starobinsky:1980te} is currently the leading paradigm of the very early universe cosmology. The existence of an inflationary period reduces the spatial curvature and increases the particle horizon, which solve the flatness problem and the horizon problem. Most importantly, the inflationary paradigm provides a predictive theory of the primordial cosmological perturbations \cite{Mukhanov:1981xt,Press:1980zz,Sato:1980yn, Hawking:1982cz,Starobinsky:1982ee,Guth:1982ec,Bardeen:1983qw}, generating a near scale-invariant power spectrum, as later observed by cosmological experiments \cite{Smoot:1992td, Aghanim:2018eyx,Akrami:2018odb}.
	 
	Despite its success, there are many open questions (sometimes considered as problems) for inflationary cosmology, see \cite{Brandenberger:2009jq, Brandenberger:1999sw,Brandenberger:2002wm} for reviews. The inflationary spacetime is past-incomplete \cite{Borde:2001nh}, and may have an initial singularity \cite{Borde:1993xh}. This suggests that inflation cannot provide a complete description of the very early universe. Another issue is the trans-Planckian problem for cosmological fluctuations \cite{Martin:2000xs}. The inflationary fluctuations may potentially begin in a super-Planckian region where currently known physical laws may be invalid. A further issue is that, for inflation to happen, the pre-inflationary universe is restricted to be somewhat old, large and homogeneous \cite{Linde:2005ht, Goldwirth:1991rj}, introducing new initial condition problems. In some inflation models, e.g. the chaotic inflation \cite{Linde:2005ht} and eternal inflation \cite{Linde:1993xx}, this prerequisite can be generic, but a measure problem arises relating to whether we are typical observers \cite{Page:2006ys, Hartle:2007zv, Page:2007bt}. Also, models of inflationary cosmology relying on scalar fields need to satisfy stringent constraints, which constitutes another initial condition problem for inflation \cite{Adams:1990pn}. 

	These open questions do not imply a failure of the inflationary paradigm. In fact, many possible solutions have been proposed within the inflationary paradigm for each of the problems here. However, we need to look out of the paradigm of inflation, to judge whether these solutions are economical or artificial. Thus, to measure the success of inflation, it is important to explore possible alternative scenarios to inflation. And these alternative theories may have a chance to succeed in their own rights.

	Generally, an alternative to inflation scenario is expected to generate near scale-invariant fluctuations, as these fluctuations are already observed. Usually, this automatically solves the horizon problem. At best, the scenario should also solve the flatness problem and possibly other problems that inflation encounters.

	There have been many alternative to inflation scenarios in the literature. To the best of our knowledge, all of these alternatives can be classified by the time-dependence of the scale factor where the time variable is real. Some examples, sorted not by physical realization but by the scale factor behavior in real time, include fast contraction \cite{Wands:1998yp, Finelli:2001sr}, slow contraction \cite{Khoury:2001wf}, slow expansion \cite{Piao:2003ty, Nayeri:2005ck, Brandenberger:2006xi, Wang:2012bq} and fast non-exponential expansion \cite{Mukohyama:2009gg}. For a review of alternative scenarios to inflation, see \cite{Brandenberger:2009jq}.

	In this paper, we propose a different approach. The scale-invariant fluctuations are generated inside an instaton, where the time variable is complex. In the Lorentzian point of view, the universe emerges quantum mechanically with all the fluctuations already present at the emergence of the universe. 
	
	To describe the quantum creation of the universe, we presume the formalism of Euclidean quantum gravity (EQG) with the no boundary initial condition \cite{Hartle:1983ai,Hawking:1983hj,Halliwell:1984eu, Hawking:1980gf, Hawking:1978jn,Gibbons:1994cg} (see also \cite{Linde:1983mx,Vilenkin:1984wp} for a different approach). In this formalism, the universe is created ``from nothing'' (i.e. from a state with scale factor $a=0$) quantum mechanically. The relative probability amplitude of the quantum creation is determined by a Euclidean path integral. In the minimal setup, the quantum created universe in EQG is curvature dominated or empty \cite{Hawking:1984hk}. And the quantum fluctuations are in their Bunch-Davies vacuum states after the quantum creation \cite{Halliwell:1984eu}. Such universes cannot be considered as alternatives to inflation since they are not suitable in size, and have no scale-invariant fluctuations in it.

	Recently, \cite{Bramberger:2017tid} proposed that in Horava-Lifshitz (HL) gravity \cite{Horava:2009if}, the universe starts its cosmological evolution in a relatively flat configuration and the flatness problem is resolved. A large universe can be quantum created if suitable model parameters are chosen. 
	In this paper, we will use the background model in \cite{Bramberger:2017tid} to create a large and flat universe. After discussing the flatness problem in detail, we work out the scalar perturbation and show that the near scale-invariant power spectrum can be obtained from a Lifshitz scalar field. 

	The paper is organized as follows. In section \ref{sec:model} we introduce our model. Then in section \ref{sec:background} we analyze the dynamics of cosmological background as well as the flatness problem. After that, we study the scalar perturbation and discuss how the observed near scale-invariant power spectrum can be obtained in section \ref{sec:perturbation}. A numerical study is then given in section \ref{sec:num} to verify some approximations in previous sections. Finally, we conclude in section \ref{sec:conclusion} and discuss some future directions. 
	
	\section{A Concrete Model}
	\label{sec:model}
	\subsection{The Metric and the Action}
	We decompose the metric using the ADM formalism \cite{Arnowitt:1959ah}:
	\begin{equation}
	ds^2 = -N^2dt^2 + h_{ij}(dx^i + N^idt)(dx^j + N^jdt)~.
	\end{equation}
	Here $t$ is the time coordinate, $x^i$ ($i$=1, 2, 3) are spatial coordinates. The constraints and dynamics in the metric are encoded in the lapse function $N(t)$, shift vector $N^i(t,\vec{x})$ and the 3-metric $h_{ij}(t,\vec{x})$.
	
	We consider the projectable version of HL gravity (see appendix \ref{app:HL} for a brief introduction). In 4 spacetime dimensions, the most general action for $z=3$ HL theory minimally coupled with a scalar field is 
	\begin{equation}
		I = I_g + I_m \label{action}~,
	\end{equation}
	where $I_g$ and $I_m$ are restricted by the HL symmetry and their most general forms are
	\begin{equation}
	I_g = \frac{1}{2} \int Ndtd^3x \sqrt{h}(K_{ij}K^{ij} - \lambda K^2 - 2\Lambda + R + L_{z=3}) \label{Ig} ~,
	\end{equation}
	where
	\begin{equation}
	L_{z=3} = c_1 D_iR_{jk}D^iR^{jk} + c_2 D_iRD^iR + c_3R_i^jR_j^kR_k^i + c_4RR_i^jR_j^i + c_5R^3~.
	\end{equation} 
	The matter action takes the form \cite{Chen:2009ka}
	\begin{equation}
		I_m =\int Ndtd^3x \frac{1}{2}\sqrt{h}\left[ \frac{1}{N^2} (\partial_t \phi - N^i \partial_i \phi)^2 - \sum_{J \geq 2} \sum_{n=0}^3 (-1)^n \frac{\lambda_{J,n}}{M^{2n+J-4}}\Delta^n \star \phi^J \right] \label{Im} ~.
	\end{equation}

	\subsection{Cosmological Evolution Indicated by EQG Formalism}
    In EQG formalism, the universe starts in the quantum gravity (QG) region (or called the Euclidean region since the time is imaginary), during which the state of the universe is described by its wave function $\Psi[h_{ij},\phi]$, defined as the path integral over possible compact configurations with boundary geometry and matter be $h_{ij}$ and $\phi$:
    \begin{equation}
    \Psi[h_{ij},\phi] = \int d[g_{\mu \nu}] d[\phi] e^{-\hat{I}} ~,\label{Psi}
    \end{equation}
	where $h_{ij}$ and $\phi$ are the three-geometry and matter field configuration at the end of the state, and $\hat{I}$ is the Euclidean action obtained from \eqref{action} by a Wick rotation of time $\tau = i t$. Hawking's no boundary proposal \cite{Hawking:1981gb} is applied, so the path integral \eqref{Psi} is defined over all possible paths which start from a vanishing universe, go through compact configurations and end in the final configuration $[h_{ij},\phi]$.
	
	The relative probability for the universe to be in a certain $[h_{ij},\phi]$ configuration is $\mathcal{P}[h_{ij},\phi] \propto |\Psi [h_{ij},\phi] |^2$. We assume that the dynamics of the universe is described by the most probable path, that is, the path that follows the classical equation of motion (EoM) with background geometry to be the $k=1$ Friedmann-Robertson-Walker (FRW) metric \cite{Wiltshire:1995vk}.  
	
	The universe will end its QG evolution and tunnel to Lorentzian spacetime, and the time becomes real.  After the quantum tunneling, the evolution of the universe is predicted by standard cosmology. An inverse Wick rotation is presented: $t = t_T + \int (-id\tau)$ where $t_T$ represents the Lorentzian time at the quantum tunneling event.
	
	\subsection{Spectator Field Approximation}
	For simplicity, we use spectator field approximation throughout this work. By ``spectator'', gravity is just a background and does not receive back-reaction from the scalar field. The field fluctuation may convert to curvature perturbation through a rolling background, or the curvaton mechanism \cite{Enqvist:2001zp, Lyth:2001nq} or the modulated reheating \cite{Dvali:2003em, Kofman:2003nx, Suyama:2007bg}.
	
	The spectator field approximation allows us to separate the gravity action $I_g$ and matter action $I_m$. The wave function of universe \eqref{Psi} can be decomposed into
	\begin{equation}
		\Psi[h_{ij},\psi] = \Psi_g[h_{ij}] \Psi_m[\phi], \Psi_g[h_{ij}] \equiv \int d[h_{ij}]e^{-\hat{I}_g}, \Psi_m[h_{ij},\phi] \equiv \int d[\phi]e^{-\hat{I}_m}~.
	\end{equation}
	The gravity part $\Psi_g[h_{ij}]$ uniquely determines the dynamics of the background. Under the spectator field approximation, there is no metric perturbation, so the scalar perturbation is uniquely determined by $\Psi_m[h_{ij},\phi]$.
	
	\section{The Background Cosmology}	\label{sec:background}		
	\subsection{Classical Background EoM}
	We consider a homogeneous and isotropic closed universe, the metric is:
	\begin{equation}
	ds^2 = -dt^2 + a^2(t) \left(\frac{d\bar{r}^2}{1-\bar{r}^2} + \bar{r}^2 d\Omega^2 \right)~.
	\end{equation}
	Here $d\bar{r}^2/(1-\bar{r}^2) + \bar{r}^2 d\Omega^2$ is the metric of a unit sphere, so the scale factor $a(t)$ represents the radius of the universe and has the dimension of length.
	
	The variation of \eqref{Ig} with respect to $a(t)$ gives the classical EoM
	\begin{equation}
	\frac{3\lambda-1}{2}(2\partial_t H + 3H^2) = \frac{\gamma }{a^6} - \frac{1}{a^2} + \Lambda ~,
	\end{equation}
	where $H \equiv (\partial_t a)/a$ and $\gamma \equiv 24(c_3+3c_4+9c_5)$. Integrating the equation, we get
	\begin{equation}
	\frac{3}{2}(3\lambda - 1) H^2 = \frac{C}{a^3} - \frac{\gamma}{a^6} - \frac{3}{a^2} + \Lambda \label{Lorentzbgeom} ~,
	\end{equation}
	where $C$ is the integration constant which behaves as dark matter \cite{Mukohyama:2009mz}. A discussion on the constraint on $C$ is included in appendix \ref{app:Cnon0}.
	
	\subsection{Background Dynamics in the Euclidean Region}
	In the Euclidean region, the time is imaginary. We make a Wick rotation $\tau = it$, and define $\mathcal{H} \equiv \frac{\partial_{\tau} a}{a} = -iH$. Also for convenience, we define $\tilde{\lambda} \equiv \frac{3}{2}(3\lambda - 1)$. In the IR region, our model should recover standard GR in which $\lambda = 1$ due to higher symmetry. Since there exists an IR Gaussian fixed point where $\lambda = 1$ and $\Lambda = 0$ \cite{Contillo:2013fua}, we may reasonably take $\tilde{\lambda} > 0$ and $\Lambda = 0$. The EoM \eqref{Lorentzbgeom} is then:
	\begin{equation}
	\tilde{\lambda} \mathcal{H}^2 = - \frac{C}{a^3} + \frac{\gamma}{a^6} + \frac{3}{a^2} \label{Eubgeom} ~.
	\end{equation}
	Here $\tilde{\lambda}$ is dimensionless, $\mathcal{H}$ has dimension $[T]^{-1}$ or $[L]^{-1}$, and $a$ has dimension $[L]$. So the model parameter $C$ and $\gamma$ should have dimension $[L]$ and $[L]^4$, respectively. 
	
	\subsection{Initial Condition and Final State of Euclidean Evolution}
	In EQG, the universe starts with a vanishing size $a=0$ in the Euclidean regime. In later calculations, we need the asymptotic behavior of $a(\tau)$ near $\tau = 0$. Expanding $a(\tau)$ in the neighbor of $\tau = 0$ and keeping the two leading terms, we have
	\begin{equation}
	a(\tau) \sim a_1 \tau^{n_1} + a_2 \tau^{n_2},\quad
	a_1, a_2 \neq 0, \quad 
	n_2 > n_1 > 0, \quad
	\tau \to 0 \label{aexpand} ~.
	\end{equation}
	Substituting \eqref{aexpand} into \eqref{Eubgeom} and equating the two highest power of $\tau$ gives
	\begin{equation}
	a(\tau) \sim a_1 \tau^{\frac{1}{3}} + a_2 \tau^{\frac{4}{3}} = \left( \frac{9\gamma}{\tilde{\lambda}} \right)^{\frac{1}{6}} \tau^{\frac{1}{3}} - \frac{C}{4}\left( \frac{3}{\gamma \tilde{\lambda}^2} \right)^{\frac{1}{3}} \tau^{\frac{4}{3}}, \quad
	\tau \to 0 \label{bginitial}~.
	\end{equation}
	
	At the end of the Euclidean evolution, the universe tunnels to Lorentzian spacetime, and the time becomes real. Labeling the tunneling time as $\tau_T$, the condition that quantum tunneling happens is $(\partial_{\tau} a)_{|\tau = \tau_T} = 0$, or equivalently $\mathcal{H}(\tau_T) = 0$.
	
	Define $a_T \equiv a(\tau_T)$, the tunneling condition along with \eqref{Eubgeom} gives: $C a_T^3 = \gamma + 3a_T^4$. To generate a flat enough universe, the curvature component at the moment of tunneling $\frac{3}{a_T^2}$ should be subdominant compared to other terms: $\frac{3}{a_T^2} \ll \min (\frac{C}{a_T^3},\frac{\gamma}{a_T^6})$. This gives an approximated value $a_T \approx (\gamma/C)^{\frac{1}{3}}$.
	
	\subsection{The Flatness of the Universe}
	The conditions required to create a universe flat enough to be consistent with observations is that the model should predict a subdominant curvature component at the very beginning of the Lorentzian region. If this was true, at $\tau_T$ we have $a_T \approx (\gamma/C)^{\frac{1}{3}}$ and 
	\begin{equation}
		\frac{3/a_T^2}{C/a_T^3} \ll 1 \to \frac{\gamma}{C^4} \ll 1 \label{flat1} ~.
	\end{equation}
	Since $\gamma$ is a non-zero quantity, \eqref{flat1} implies that a large $C$, or equivalently, a strong presence of dark matter is needed for generating a flat enough universe. 
	
	Physically, the HL coupling dominates at first for its $a^{-6}$ dependence. As the scale factor increases, the curvature term becomes more and more important, so a relatively large $C$ can help to end the Euclidean evolution quickly to avoid the domination of the curvature component. Thus, the flatness problem is turned into a fine-tuning problem between the model parameter and an integration constant\footnote{This fine-tuning problem is not a technical naturalness problem since $C$ is an integration constant, similar to the situation of unimodular gravity with the cosmological constant problem \cite{Weinberg:1988cp}.}. 
		
	\section{Analysis of Perturbations}	\label{sec:perturbation}
	\subsection{Quadratic Action of Perturbations}
	Recall that, the perturbations come solely from the scalar field under the spectator field approximation, of which the most general action is \eqref{Im}; if we substitute $\sqrt{h}$, $N$ and $N^i$ to its background value $a^3$, 1 and $0$, and take only $J=2$ terms for simplicity, the action becomes:
	\begin{equation}
	I_m =\int dtd^3x \frac{1}{2}a^3\left[ (\partial_t \phi)^2 -  \sum_{n=0}^3 (-1)^n \frac{\lambda_{2,n}}{M^{2n-2}}\Delta^n \star \phi^2 \right] \label{Imwithbg} ~.
	\end{equation}
	
	We separate the matter field $\phi$ as a homogeneous and isotropic background $\bar{\phi}$ and perturbation $\delta \phi$, which can be expanded as a summation of spherical harmonics:
	\begin{equation}
	\phi(t,\vec{x}) = \bar{\phi}(t) + \delta \phi(t,\vec{x}) = \bar{\phi}(t) + \sum_k f_k(t)Q_k(\vec{x}) \label{phidef} ~,
	\end{equation}
	where $Q_k(\vec{x})$ is defined as the normalized eigenfunction of the Laplacian operator on three-dimensional unit sphere $S^3$: $\Delta Q^k = -(k^2-1)Q^k$ and $\int_{S^3} d^3x Q_nQ_m = \delta_{nm}$. We have suppressed the other two indices since they are irrelevant to our work. Inserting \eqref{phidef} into \eqref{Imwithbg}, the quadratic action for perturbation modes is:
	\begin{equation}
	    I_{m,2} = \int dt \frac{a^3}{2} \sum_k \left( \dot{f}_k^2 - \omega_k^2 f_k^2 \right) ~,
	\end{equation}
	where a dot denotes differentiation with respect to time, and $\omega_k$ is defined by
	\begin{equation}
	\omega_k^2 \equiv \sum_{n=0}^3  \frac{\lambda_{2,n}}{M^{2n-2}} (k^2-1)^n \approx \sum_{n=0}^3  \frac{\lambda_{2,n}}{M^2} \left( \frac{k}{M} \right)^{2n} \label{omegadef} ~.
	\end{equation}
	The approximation comes from the fact that the wavelengths of currently observed perturbations are much larger than $1$. Also, we can obtain the Euclidean action for perturbation by a Wick rotation $\tau = it$:
	\begin{equation}
		\hat{I}_{m,2} = \int d\tau \frac{a^3}{2} \sum_k \left[ f_k^{\prime 2} + \omega_k^2 f_k^2 \right] \label{Euptaction} ~,
	\end{equation} 
	where a prime denotes differentiation with respect to $\tau$.
	
	From \eqref{Euptaction}, perturbations of different wavenumbers are separable, which enables us to define the action for a single wavenumber $k$ of perturbation:
	\begin{equation}
		\hat{I}_{m,2k} = \int d\tau \frac{a^3}{2}\left[ f_k^{\prime 2} + \omega_k^2 f_k^2 \right] \label{Im2k} ~.
	\end{equation}
	
	\subsection{Classical EoM and Initial Asymptotic Behavior of Perturbations}
	The classical EoM for $f_k$ comes from the variation of \eqref{Im2k} with respect to $f_k$:
	\begin{equation}
        f_k^{\prime \prime} + 3\mathcal{H}f_k^{\prime} - \omega_k^2 f_k = 0 \label{fkeom} ~.
	\end{equation}
	For later convenience, we shall introduce $g_k \equiv \frac{f_k^{\prime}}{f_k}$. Then \eqref{fkeom} becomes
	\begin{equation}
	g_k^{\prime} + g_k^2 + 3\mathcal{H}g_k - \omega_k^2 = 0 \label{gkeom} ~.
	\end{equation}
		
	The asymptotic behavior of $a(\tau)$ around $\tau = 0$ can uniquely determine the asymptotic behavior of perturbations. Recall that $a(\tau) \sim a_1 \tau^{\frac{1}{3}}$ near $\tau = 0$, this gives $\mathcal{H}(\tau) \sim \frac{1}{3\tau}$, and \eqref{fkeom} becomes
	\begin{equation}
		f_k^{\prime \prime} + \frac{3}{\tau} f_k^{\prime} - \omega_k^2 f_k = 0, \quad \tau \to 0 \label{fkeom0} ~.
	\end{equation}	
	\eqref{fkeom0} has the form of a Bessel equation, and its general solution is
	\begin{equation}
	f_k = c_1 J_0(i\omega_k \tau) + c_2 N_0(i\omega_k \tau) ~,
	\end{equation}	
	where $J_{\nu}(z)$ and $N_{\nu}(z)$ are the Bessel and Neumann functions. Note that $N_0(z)$ is divergent near $z=0$, so $c_2$ should be set to 0 for a finite perturbation. Furthermore, as we shall see later, the magnitude of $f_k$ is irrelevant to our work, so we can safely take $c_1 = 1$ and the only possible asymptotic form of $f_k$ is
	\begin{equation}
	f_k(\tau) = J_0(i \omega_k \tau), \quad \tau \to 0 \label{fkini} ~.
	\end{equation}	
	The asymptotic behavior of $g_k$ around $\tau = 0$ is
	\begin{equation}
		g_k(\tau) = -i\omega_k \frac{J_1(i\omega_k \tau)}{J_0(i \omega_k \tau)}, \quad \tau \to 0 \label{gkini} ~.
	\end{equation}	
	
	\subsection{The Wave Function of Perturbation Modes}
    The decomposition of perturbations with different wavenumbers enables us to define the wave function for $f_k$:
	\begin{equation}
		\Psi_m[a,f_k] \equiv \int d[f_k]e^{-\hat{I}_{m,2k}} \label{Psifk} ~.
	\end{equation}
	
	The path integral \eqref{Psifk} can be evaluated by the WKB approximation, which states that the dominating contribution of the path integral comes from the neighbor of classical trajectory. Following the method in \cite{Halliwell:1984eu}, we use integration by parts to transform \eqref{Im2k} to
	\begin{equation}
		\hat{I}_{m,2k} = \left( \int d\tau f_k D^S_k f_k \right) + \hat{I}_{B,k} ~,
	\end{equation}
	where $D^S_k$ is the operator satisfying the classical equation of motion $D^S_k f_k = 0$, which originates from the variation of \eqref{Im2k} with respect to $f_k$. Explicitly we have
	\begin{equation}
		D^S_k = -\frac{1}{2}a^3 \left( \frac{\partial^2}{\partial \tau^2} - \omega_k^2 + 3 \mathcal{H} \frac{\partial}{\partial \tau} \right) ~,
	\end{equation}	
	and $\hat{I}_{B,k}$	is the induced boundary term
	\begin{equation}
		\hat{I}_{B,k} = \left( \frac{1}{2}a^3 f_k \partial_{\tau} f_k \right) _{|^{\tau_T}_0} = \left( \frac{1}{2}a^3 f_k \partial_{\tau} f_k \right)_{\tau_T} ~.
	\end{equation}
	
	Under the WKB approximation, $\left( \int d\tau f_k D^S_k f_k \right)$ does not contribute to the correlation function. The main contribution to the two-point correlation function comes from the quadratic term of $\hat{I}_{B,k}$:
	\begin{equation}
		\hat{I}_{B,k2} \sim \frac{1}{2}a(\tau_T)^3 g_k(\tau_T) f_k^2 ~,
	\end{equation}
	and the wave function is rewritten as
	\begin{equation}
		\Psi_m[f_k] = \int d[f_k]e^{\frac{1}{2}a(\tau_T)^3 g_k(\tau_T) f_k^2} ~.
	\end{equation}
		
	\subsection{Two-Point Correlation Function and Power Spectrum}	
	The two-point correlation function of $f_k$ at $\tau_T$ can be evaluated as
		\begin{equation}
		\langle f_kf_k \rangle = \frac{\int [df_k] f_kf_k e^{-\hat{I}_{m,2k}}}{\int [df_k] e^{-\hat{I}_{m,2k}}} = a(\tau_T)^{-3}g_k^{-1}(\tau_T) \label{2pt} ~.
		\end{equation}
	Note that, \eqref{2pt} is independent on the overall magnitude of $f_k$, which justifies the specific choice of $f_k$ in \eqref{fkini}. Besides, \eqref{2pt} implicitly contains a delta function due to the decomposition of different wavenumbers.
	
	\eqref{2pt} is implicitly related to the model parameters by the classical EoM of $g_k$ \eqref{gkeom}. As we will show numerically in section \ref{sec:num}, the solution to \eqref{gkeom} near the tunneling event $\tau = \tau_T$ can be approximated as $g_k(\tau) \simeq \omega_k$. We may understand this fact by the observation that, the $3\mathcal{H} g_k$ term is subdominant close to $\tau = \tau_T$ from the tunneling condition $\mathcal{H}(\tau_T) = 0$, and the remaining equation $g_k^{\prime} + g_k^2 = \omega_k^2$ has a general solution $g_k(\tau) = \omega_k \tanh[\omega_k(\tau + \tau_c)]$, which approximates to $\omega_k$ at large enough $\tau$. \eqref{2pt} can then be explicitly expressed by the model parameters:
	\begin{equation}
		\langle f_kf_k \rangle = a(\tau_T)^{-3}\omega_k^{-1} = \frac{C}{\gamma \omega_k} \label{2ptsimp} ~.
	\end{equation}
	Now we see that the $k$ dependence of the two-point correlation function comes solely from $\omega_k$, since $C$ and $\gamma$ are constants. Comparing with the definition of the power spectrum
	\begin{equation}
	\langle f_{k_1}f_{k_2} \rangle \equiv (2\pi)^3 \delta(\vec{k_1}+\vec{k_2}) \frac{2\pi^2}{k^3} P_{\phi}(k) \label{psdef} ~,
	\end{equation}
	we have
	\begin{equation}
	\omega_k \simeq \frac{k^3}{(2\pi)^5 a^3(\tau_T)} \frac{1}{P_{\phi}(k)} ~.
	\end{equation}
	To generate a near scale-invariant power spectrum (i.e. $\frac{d\ln P_{\phi}(k)}{d\ln k} \sim 0$), we need $\omega_k \sim k^3$. From \eqref{omegadef} we see this can be accomplished as long as the positive $\frac{\lambda_{2,3}k^6}{M^8}$ term dominates the $\omega_k^2$ expression.
	
	\subsection{Spectral Index}
	Although the scalar perturbations are nearly scale invariant, a small tilt is observed. The spectral index as $n_s - 1 \equiv \frac{d\ln P_{\zeta}(k)}{d\ln k}$ is observationally constrained by $n_s = 0.9665 \pm 0.0038$ \cite{Akrami:2018odb}. The standard cosmology connects observed cosmological fluctuations today with those at very early times by the adiabatic modes which is conserved outside the horizon $aH > k$ \cite{Weinberg:2003sw}. In inflationary cosmology, the related perturbation modes will all exit the horizon during the quasi de-Sitter period. 
	
	In our model, however, the scalar perturbation is inside the horizon at $t_T$ because of the tunneling condition $H(t_T) = 0$, so the two-point function also depends on the physics after the tunneling. We analyze two cases and see how a slight tilt is generated.
	
	If a quick reheating process happens right after the quantum tunneling and turns the dark matter $C$ and HL coupling $\gamma$ into radiation, then the time of horizon exiting $t_{k \ast}$ is just the tunneling time $t_T$, due to the finiteness of $H$ and largeness of $a$.In this case, \eqref{2ptsimp} is valid and the tilt generated by subdominated terms in \eqref{omegadef} is
	\begin{equation}
		\frac{d\ln P_{\phi}(k)}{d\ln k} = - \frac{d\ln}{d\ln k^2}\left[ 1 + \sum_{n=0}^2 \frac{\lambda_{2,n}}{\lambda_{2,3}} \left( \frac{k}{M}\right)^{2n-6} \right] \approx \frac{M^6}{\lambda_{2,3}} \left( 3\lambda_{2,0} +  \frac{2\lambda_{2,1}}{M^2} + \frac{\lambda_{2,2}}{M^4} \right) ~,
	\end{equation}
	which had better be negative inspired by observations \cite{Akrami:2018odb}. However, in the IR region, the standard dispersion relation $\omega_k^2 = m^2 + c_s^2 k^2$ suggests that $\lambda_{2,0}$ and $\lambda_{2,1}$ should be positive. So in this case, the condition for a power spectrum with red tilt is
	\begin{equation}
		\lambda_{2,2} < 0 \quad \& \& \quad \frac{\lambda_{2,3}}{M^6} \gg \frac{|\lambda_{2,2}|}{M^4} \gg \max \left( \frac{\lambda_{2,1}}{M^2}, \lambda_{2,0} \right)  ~. \label{tiltconst}
	\end{equation}

	If the reheating process is slow, the perturbation modes will exit the horizon when the dark matter $C$ and HL coupling $\gamma$ still dominate, during which the dynamics of the universe is described by
	\begin{equation}
		\tilde{\lambda} H^2 = \frac{C}{a^3} - \frac{\gamma}{a^6} - \frac{3}{a^2} \approx \frac{C}{a^3} - \frac{\gamma}{a^6} \label{LorentzEoM}~.
	\end{equation}
	Define the time for horizon exiting of $k$ mode as $t_k$, the horizon exiting condition is $a(t_k)H(t_k) = k$, or $\tilde{\lambda} k^2 = \frac{C}{a(t_k)} - \frac{\gamma}{a(t_k)^4}$, from which we can estimate
	\begin{equation}
		a(t_k) \approx \left( \frac{\gamma}{C} \right)^{\frac{1}{3}} \left[ 1 + \frac{\tilde{\lambda}}{3}\left( \frac{\gamma}{C^4} \right)^{\frac{1}{3}}k^2 + \mathcal{O}(k^4) \right] ~,
	\end{equation}
	here the condition \eqref{flat1} is applied. Now the two-point correlation function of $f_k$ at horizon exiting is
    \begin{equation}
    	\langle f_kf_k \rangle = a(t_k)^{-3}\omega_k^{-1} = \frac{C}{\gamma \omega_k}  \left[ 1 + \tilde{\lambda} \left( \frac{\gamma}{C^4} \right)^{\frac{1}{3}}k^2 + \mathcal{O}(k^4) \right]^{-1} \label{2ptfkcomplex} ~.
    \end{equation}
	We can see that the $k^6$ term in $\omega_k$ and the $k^2$ term square brackets combined produce a $k^{-8}$ dependence in the two-point function. The red tilt can be generated from the $k^{-8}$ term, so in this case \eqref{tiltconst} is no longer required. 
	
	Finally, recall that we need a rolling background, or curvaton, or modulated reheating to convert the spectator fluctuation into curvature fluctuation. In this conversion process, a further tilt may be generated and contribute to the final spectral index.

	\section{A Numerical Study} \label{sec:num}
	We present here a numerical case. To avoid the appearance of large numbers, we rescale the parameters: we set a length scale $a_0$ with dimension $[L]$, and define
	\begin{equation}
	\tilde{a} \equiv \frac{a}{a_0}, \quad
	\tilde{\gamma} \equiv \frac{\gamma}{a_0^4}, \quad
	\tilde{C} \equiv \frac{C}{a_0}, \quad
	\mathcal{H} = \frac{\partial_{\tau} a}{a} = \frac{\partial_{\tau} \tilde{a}}{\tilde{a}} ~.
	\end{equation}
	Now $\tilde{a}$, $\tilde{\lambda}$, $\tilde{\gamma}$ and $\tilde{C}$ are all dimensionless. The EoM \eqref{Eubgeom} then becomes
	\begin{equation}
	\tilde{\lambda} a_0^2 \mathcal{H}^2 = - \frac{\tilde{C}}{\tilde{a}^3} + \frac{\tilde{\gamma}} {\tilde{a}^6} + \frac{3}{\tilde{a}^2} \label{Eubgeomscaled}~.
	\end{equation} 
	
	We take $\tilde{\gamma} = 2.0 \times 10^{-2}$, $\tilde{C} = 5.0 \times 10^2$, $\tilde{\lambda} = 1.0$ and $a_0 = 1.0 \times 10^{35} l_p$. The original model parameters are then $C = 5.0 \times 10^{37} l_p$, $\gamma = 2.0 \times 10^{138}l_p^4$ and $\lambda = 0.56$. The tunneling condition $\mathcal{H}(\tau_T) = 0$ gives the constraint on scaled scale factor at the tunneling time $a_T \equiv a(\tau_T)$:
	\begin{equation}
	- \frac{\tilde{C}}{\tilde{a}^3_T} + \frac{\tilde{\gamma}} {\tilde{a}^6_T} + \frac{3}{\tilde{a}^2_T} = 0 ~.
	\end{equation}
	Numerically it gives $\tilde{a}_T = 7.4 \times 10^{-2}$. The real scale factor $a(\tau_T) = \tilde{a}_T a_0 = 7.4 \times 10^{33} l_p$. Note that $a(\tau_T)$ needs to be extremely large to be consistent with a currently flat universe. This statement is checked in Appendix \ref{app:a0}.
	
	We check how \eqref{flat1} is satisfied in this numerical case. The ratio $\frac{3/a_T^2}{C/a_T^3} = 4.4 \times 10^{-3}$ in our numerical case is small and close to the current proportion of gravity $\Omega_K = 0.0007 \pm 0.0019$ \cite{Aghanim:2018eyx}. We then see that the flatness problem is transferred to a parameter choice problem in our model. 
	
\begin{figure}[tbp]
    	\includegraphics[width=.3\textwidth]{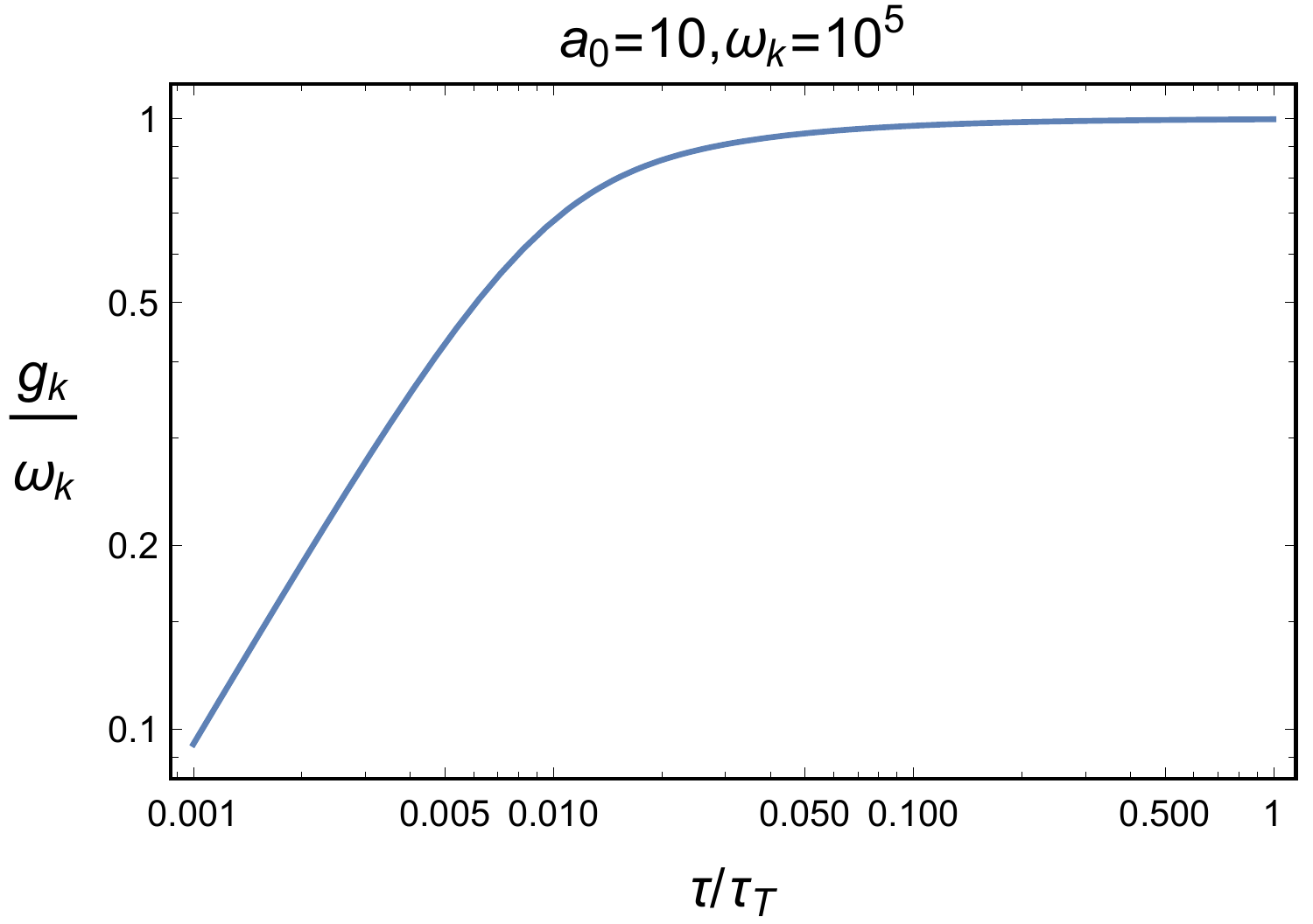}
    	\includegraphics[width=.3\textwidth,height=.1457\textheight]{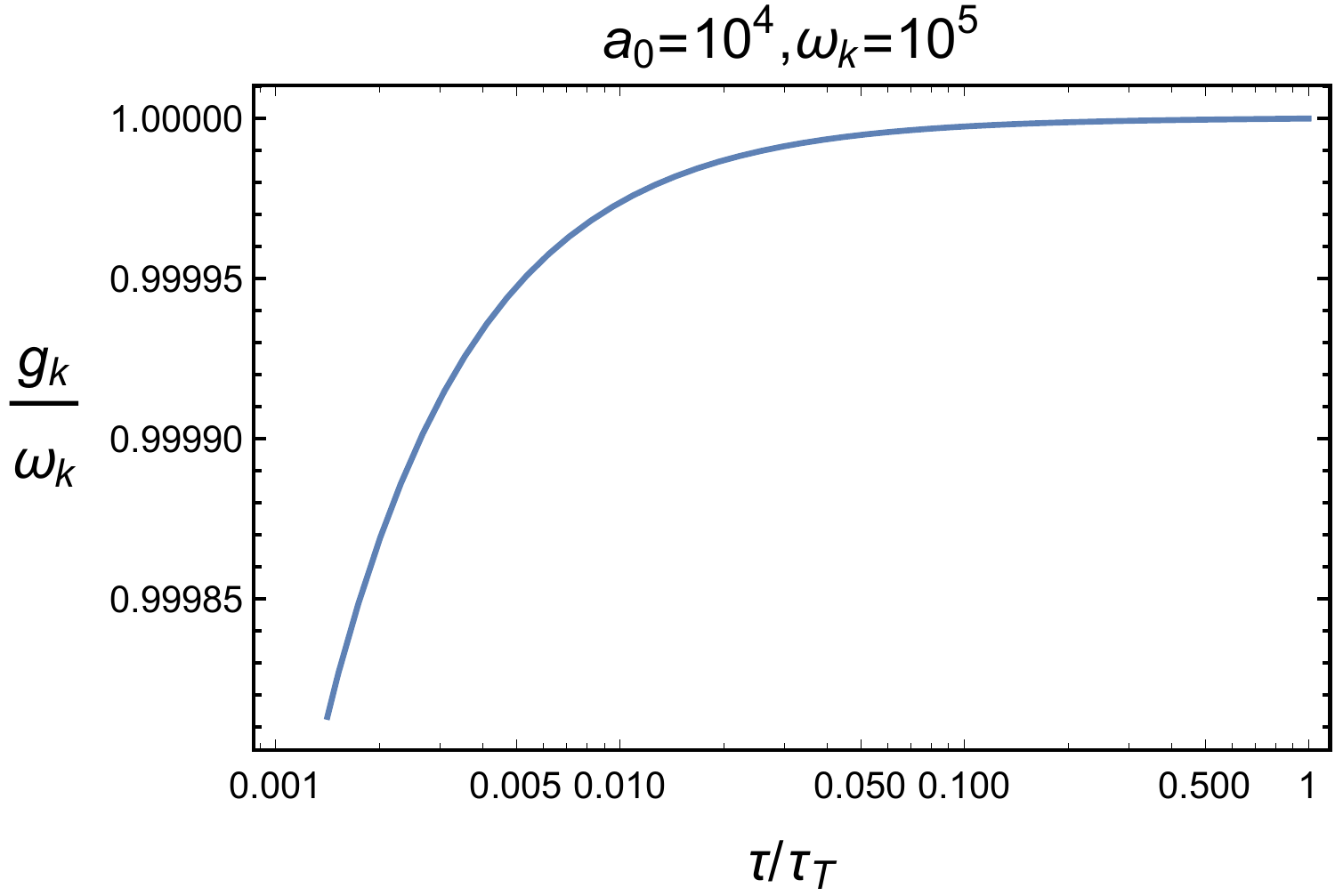}
		\includegraphics[width=.3\textwidth,height=.1442\textheight]{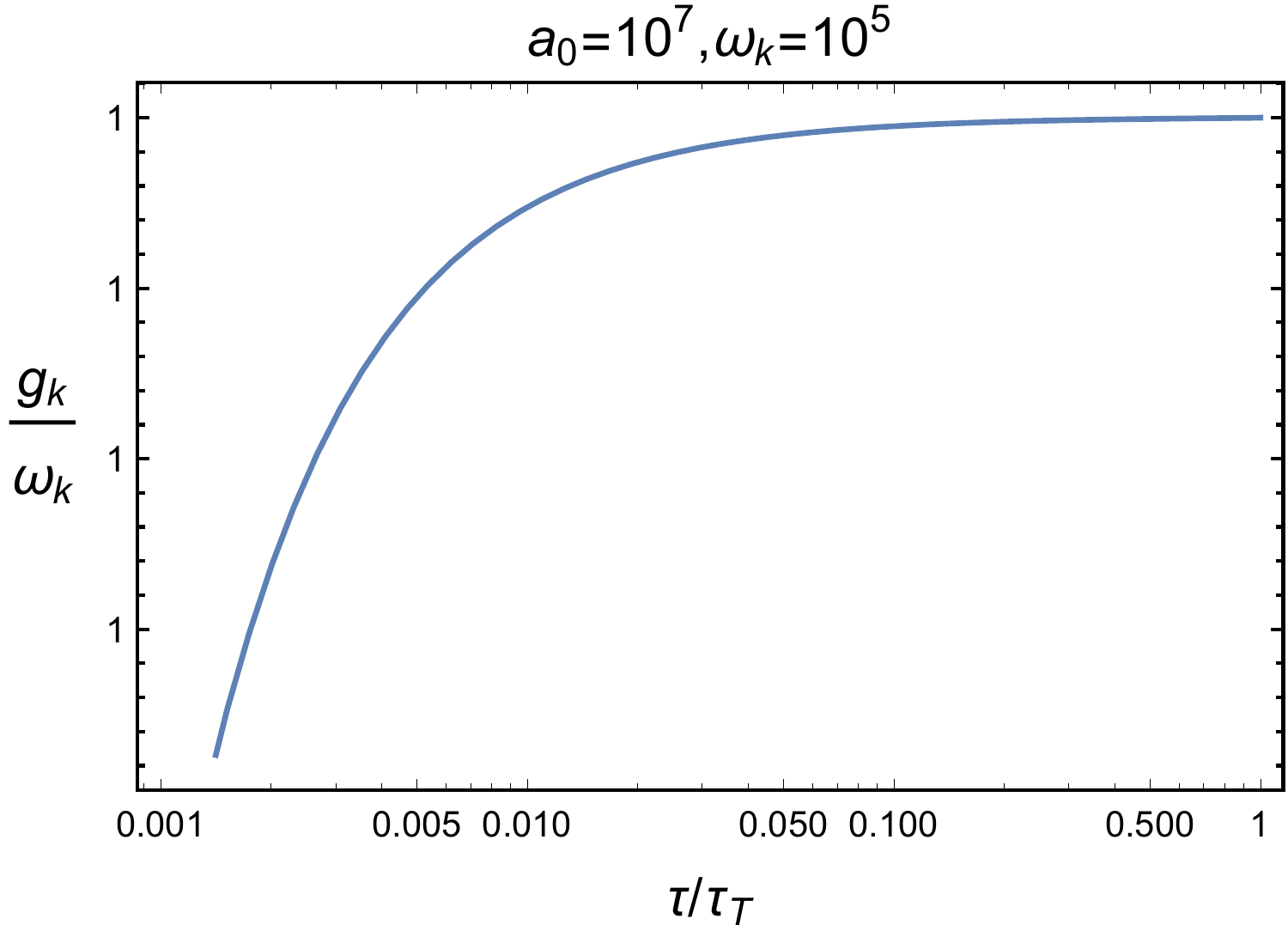}
		\caption{The $g_k/\omega_k$ ratio as a function of time. At late Euclidean time this ratio approaches to 1 for different choices of model parameters. This verifies our approximation of taking $g_k \sim \omega_k$. \label{fig:approx}}
	\end{figure}

	For perturbations, as mentioned in section \ref{sec:perturbation}, the observational effect of perturbation originates from $g_k(\tau_T)$, which is determined by \eqref{gkeom}. To show that the validity of our approximation $g_k \sim \omega_k$ in section \ref{sec:perturbation} is generic and to avoid the possible numerical deviation from large numbers, we numerically calculated $g_k(\tau)$ with different but relatively small $a_0$. The result is presented in figure \ref{fig:approx}, we see that the asymptotical behavior of $g_k \sim \omega_k$ near the tunneling point supports \eqref{gkeom}.

	\section{Conclusion and Discussion} \label{sec:conclusion}
	In this paper, we construct a toy model for alternative to inflation. The model is motivated by some current puzzles for inflationary cosmology. If we do not introduce scenarios like bouncing cosmology \cite{Gasperini:1992em,Khoury:2001wf,Novello:2008ra,Brandenberger:2012zb,Lehners:2008vx,Brandenberger:2011gk,Cai:2007qw,Cai:2008qb,Cai:2012va}, the very early universe will undergo  quantum gravity effects, especially in the initial singularity of Big Bang cosmology. In this case, we should include some theory of quantum gravity for a complete description of the very early universe. We choose the underlying theory to be Hawking's EQG formalism, since it predicts the relative probabilities of any configuration of the universe, thus the fine-tuning problems from flatness problem and initial condition problem can be transferred into a matter of probability. We have seen in section \ref{sec:background} that, with properly chosen parameters, the most probable universe created by the HH instanton will have a large radius, which is consistent with the smallness of $\Omega_K$ today. Also, the Euclidean action reaches its minimum when the universe is in the maximum symmetric configuration \cite{Coleman:1985rnk}, in EQG this means an initially homogeneous universe is the most probable, which might be employed to explain the initial condition problem.
	
	After that, we need the action of gravity and matter, as well as the boundary condition to determine the cosmological evolution. The boundary condition is chosen to be Hawking's no boundary proposal since it can interpret the initial singularity $a(0)=0$ \cite{Hawking:1981gb}. The action should be renormalizable, otherwise, the theory cannot be applied to Planck scale, contradicting with our motivation for explaining the initial singularity of the universe, which provides us an additional motivation of choosing HL action. Now the model is constructed, and we have seen that the puzzles in inflationary cosmology may get explanations under our framework. 
	
	Once we construct the toy model, it is important to check whether it can reproduce the successful prediction in inflationary cosmology. The horizon problem is automatically solved since at $\tau=0$ all modes are casually connected. We verify that our model may transfer the flatness problem into a tuning problem of model parameters in section \ref{sec:background}, and confirm that a near-scale invariant power spectrum may be generated from the Lifshitz field in section \ref{sec:perturbation}.
	
	After setting up the framework of this Euclidean alternative to inflation approach in this paper, it is interesting to explore many further directions. For example,

	\begin{itemize} \renewcommand{\labelitemi}{-}
		\item We do not fully understand QG yet and EQG may or may not be the right approximation (see for example discussions in \cite{Firouzjahi:2004mx,Sarangi:2005cs, Sarangi:2006eb}). If the EQG formalism is invalid, the toy model is unreliable.
		\item We have studied a spectator field. The spectator fluctuation can be converted to curvature by mechanisms such as curvaton \cite{Enqvist:2001zp, Lyth:2001nq} or modulated reheating \cite{Dvali:2003em, Kofman:2003nx, Suyama:2007bg}. However, it is more interesting to study the possibility if the curvature perturbation is directly created in the EQG approach. Thus, it is interesting to calculate the full perturbation theory in EDG with gravitational fluctuations.
		\item The existence of $C$ requires the violation of Hamiltonian constraints which is not observed yet; and the non-trivial geometry (see appendix \ref{app:Cnon0}) is not observable. so it is interesting to seek for other mechanisms (for example, a HL action with different anisotropy scaling, or some renormalizable $f(R)$ gravity action) to replace $C$, avoiding the above requirements.
		\item The parameter $C$ appears as an integration constant here. This constant may be explained as dark matter considering how its energy density evolves with time. However, it is important to find a mechanism to either realize a radiation dominated universe from the decay of this ``dark matter'' component or add more realistic matter into the model to take the role of starting a radiation dominated universe.
		\item We have found that the feature of power spectrum can be produced by the $z=3$ Lifshitz field; and it's natural to guess if the metric perturbation of HL gravity alone can give such a feature since both of them are characterized by the $z=3$ anisotropic scaling.
		\item How to model-independently test this scenario against inflation and other alternative scenarios? Model-independent tests of primordial universe scenarios include the spectrum of primordial gravitational waves, and primordial quantum standard clocks \cite{Chen:2015lza, Chen:2016cbe, Chen:2016qce}. It is interesting to see how these criteria work in the Euclidean time regime.
	\end{itemize}
	
	We hope to address these issues in further works.

	\acknowledgments
	 
	We thank Yifu Cai, Andrew Cohen, Qianhang Ding, Xian Gao, Shinji Mukohyama, Xi Tong, Henry Tye and Siyi Zhou for the delightful discussion. This research was supported in part by GRF Grant 16304418 from the Research Grants Council of Hong Kong.

	\appendix
	\section{Conventions}
	In this paper Planck units are used, in which $c=\bar{h}=8\pi G=1$. We keep the Planck length $l_p$, time $t_p$ and energy $E_p$ reserved, but use $l_p$ and $t_p$ interchangeably. The scale factor $a(t)$ is defined by the FRW metric with positive curvature:
	\begin{equation}
		ds^2 = -dt^2 + a^2(t) \left(\frac{d\bar{r}^2}{1-\bar{r}^2} + \bar{r}^2 d\Omega^2 \right) ~.
	\end{equation}
	Here $d\bar{r}^2/(1-\bar{r}^2) + \bar{r}^2 d\Omega^2$ is the metric of a unit sphere, so the scale factor $a$ represents the radius of the universe, which has length dimension. 
	
	\section{Preliminary Introduction to Projectable HL gravity} \label{app:HL}
	In this section, we will follow \cite{Mukohyama:2010xz}. A basic property of HL gravity is the anisotropy scaling:
	\begin{equation}
	t \to b^z t, \quad \vec{x} \to b \vec{x} ~,
	\end{equation}
	where $z$ is called the dynamical critical exponent. In 4 dimensional spacetime, anisotropic scaling of $z=3$ can make the gravity renormalizable in the UV region and can also lead to a scale-invariant cosmological perturbation \cite{Mukohyama:2009gg}. In infrared(IR) region the theory should flow to $z=1$ to recover GR. So for our purpose, we concern on the HL action composed of $z=1$ and $z=3$ gravity sector, and the most general form of which is:
	\begin{equation}
	I_g = \frac{1}{2} \int Ndtd^3x \sqrt{h}(K_{ij}K^{ij} - \lambda K^2 - 2\Lambda + R + L_{z=3}) ~,
	\end{equation}
	where
	\begin{equation}
	L_{z=3} = c_1 D_iR_{jk}D^iR^{jk} + c_2 D_iRD^iR + c_3R_i^jR_j^kR_k^i + c_4RR_i^jR_j^i + c_5R^3 ~.
	\end{equation} 
	Here $h \equiv \det(h_{ij}) $, $K_{ij} \equiv (\partial_t g_{ij} - D_iN_j - D_jN_i)/(2N)$ is the extrinsic curvature, $D_i$ and $R_{ij}$ are separately the covariant derivatives and the Ricci tensor of $h_{ij}$ while $R \equiv h^{ij}R_{ij}$ is the intrinsic curvature. $\lambda$ and $c_n$(n=1,...,5) are constants.
	
	The existence of anisotropic scaling $z$ breaks the Lorentz symmetry. Instead, the fundamental symmetry of the theory is are invariance under space-independent time reparametrization and time-dependent spatial diffeomorphism:
	\begin{equation}
		t \to t^{\prime}(t), \quad  \vec{x} \to \vec{x}^{\prime}(t, \vec{x}) ~.
	\end{equation}
	The symmetry, as well as the fact that $N$ only depends on $t$, allows us to set $N=1$ by a time reparametrization $t^{\prime} = \int Ndt $.
	
	The most general renormalizable action for a single scalar field compatible in $d+1$ dimension with the symmetry of HL theory is \cite{Chen:2009ka}	
	\begin{equation}
	I_m =\int Ndtd^dx \frac{1}{2}\sqrt{h}\left[ \frac{1}{N^2} (\partial_t \phi - N^i \partial_i \phi)^2 - \sum_{J \geq 2} \sum_{n=0}^{n_J}(-1)^n \frac{\lambda_{J,n}}{M^{2n+J-4}}\Delta^n \star \phi^J \right] \label{Imfull} ~,
	\end{equation}
	where $\Delta$ is the Laplacian operator, $n_J$ is restricted by $n_J = \max \{ n \in Z | n \leq \frac{z+d}{2} + \frac{z-d}{4}J \}$, $\lambda_{J,n}$ and $M$ are constants. The $\star$ represents all possible independent combinations of $\Delta$ and $\phi$ up to a total derivative, for example:
	\begin{equation}
	\Delta^2 \star \phi^3 = a_1 \phi(\Delta \phi)^2 + a_2 \phi^2 \Delta^2 \phi ~.
	\end{equation}
	For our case, $z=3$ and $d=3$, \eqref{Imfull} becomes
	\begin{equation}
		I_m =\int Ndtd^3x \frac{1}{2}\sqrt{h}\left[ \frac{1}{N^2} (\partial_t \phi - N^i \partial_i \phi)^2 - \sum_{J \geq 2} \sum_{n=0}^3 (-1)^n \frac{\lambda_{J,n}}{M^{2n+J-4}}\Delta^n \star \phi^J \right] ~.
	\end{equation}
	
	\section{Using Non-Trivial Geometries to Release the Constraint on $C$} \label{app:Cnon0}
	In canonical gravity theory, the total Hamiltonian of gravity is 0 due to the existence of a secondary constraint (the Hamiltonian constraint) \cite{dirac2001lectures, DeWitt:1967yk}:
	\begin{equation}
	\mathcal{H}_g \equiv \frac{1}{2}\sqrt{h}(K_{ij}K^{ij} - \lambda K^2 + 2\Lambda - R - L_{z=3}) \label{HConst} = 0 ~.
	\end{equation}
	\eqref{Lorentzbgeom}, \eqref{HConst} combined give $C = 0$. To make the integration constant $C$ non-vanishing, the spatial geometry is assumed to have non-trivial topology, composed of several connected pieces, but each of which is disconnected with others\cite{Mukohyama:2009tp}. The constraint then becomes a summation 
	\begin{equation}
	\mathcal{H}_g = 0 \quad \to \quad \sum \mathcal{H}_g = 0, \qquad \qquad  
	C = 0 \quad \to \quad \sum C = 0 ~.
	\end{equation}
	Take our universe to be one of the connected pieces, the restriction of integration constant $C$ is now evaded.
	
	\section{Estimation on Scale Factor at Quantum Tunneling} \label{app:a0}
	Here we estimate the parameter space of $a(t_T)$. Given today's observation (here $t_c$ denotes the time today) \cite{Aghanim:2018eyx}
    \begin{equation}
		|\Omega_K|(t_c) < 0.0007, \quad 
		H(t_c) = 67.66 (\mathrm{km}/\mathrm{s})/\mathrm{Mpc} = 1.2 \times 10^{-61} t_p^{-1} ~,
    \end{equation}
    we can estimate a lower limit of today's radius of universe $a(t_c) > 3.2 \times 10^{62}l_p$. We then use the redshift of matter-radiation domination epoch $z(t_{eq}) = 3.4 \times 10^3$ to get $a(t_{eq}) = \frac{a(t_c)}{1+z(t_{eq})} = 9.4 \times 10^{58}l_p$.
    
    Now the energy scale at $t_{eq}$ is $\Lambda_{eq} = 0.75 \mathrm{eV}$. The energy scale at quantum tunneling $\Lambda_T$ should be larger than $100\mathrm{GeV}$, the EW phase transition scale, but less than the Planck scale $E_p = 1.2 \times 10^{19}\mathrm{GeV}$. We then have
    \begin{equation}
		a(t_T) \sim a(t_{eq}) \frac{\Lambda_{eq}}{\Lambda_T} 
		\quad \to \quad  
		a(t_T) = \left( \frac{\gamma}{C} \right)^{\frac{1}{3}} \in [5.9 \times 10^{30}, 7.0 \times 10^{47}]l_p ~.
    \end{equation}
    In section \ref{sec:num}, we choose parameters so that $a(t_T) = 7.4 \times 10^{33}l_p$. The corresponding energy scale is $\Lambda_T \sim 9.5 \times 10^{15} \mathrm{GeV}$, which is close to the GUT scale $\Lambda_\mathrm{GUT} \sim 10^{16} \mathrm{GeV}$.

\bibliography{universe}
\bibliographystyle{unsrt}
	
\end{document}